\documentclass[
showpacs,
floatfix,
aps,
prl,
twocolumn,
superscriptaddress,
amssymb,
]{revtex4-1}
\usepackage{amssymb}
\usepackage{latexsym}
\usepackage[dvips]{graphicx}
\usepackage{amsmath}
\usepackage{float}

\usepackage{graphicx}
\usepackage{dcolumn}
\usepackage{amsfonts}
\usepackage{bm}

\newcommand{\be}{\begin{equation}}
\newcommand{\ee}{\end{equation}}
\newcommand{\bea}{\begin{eqnarray}}
\newcommand{\eea}{\end{eqnarray}}

\def\bip{B_\parallel}

\def\rxx{R_{xx}}
\def\ryy{R_{yy}}

\def\easy{\left < 110 \right >}
\def\hard{\left < 1\bar{1}0 \right >}
\def\x{\hat{x}}
\def\y{\hat{y}}
\def\easy{\left < 110 \right >}
\def\hard{\left < 1\bar{1}0 \right >}

\newcommand{\rfig}[1]{Fig.\,\ref{#1}}

\newcommand{\rref}[1]{Ref.\,\onlinecite{#1}}
\newcommand{\rrefs}[2]{Refs.\,\onlinecite{#1},\,\onlinecite{#2}}

\begin{document}
\title{Apparent temperature-induced reorientation of quantum Hall stripes}
\author{Q.~Shi}
\affiliation{School of Physics and Astronomy, University of Minnesota, Minneapolis, Minnesota 55455, USA}
\author{M.~A.~Zudov}
\email[Corresponding author: ]{zudov@physics.umn.edu}
\affiliation{School of Physics and Astronomy, University of Minnesota, Minneapolis, Minnesota 55455, USA}
\author{B.~Friess}
\affiliation{Max-Planck-Institute for Solid State Research, Heisenbergstrasse 1, D-75569 Stuttgart, Germany}
\author{J.~Smet}
\affiliation{Max-Planck-Institute for Solid State Research, Heisenbergstrasse 1, D-75569 Stuttgart, Germany}
\author{J.\,D. Watson$^\#$}
\affiliation{Department of Physics and Astronomy, Purdue University, West Lafayette, Indiana 47907, USA}
\affiliation{Birck Nanotechnology Center, Purdue University, West Lafayette, Indiana 47907, USA}
\author{G.\,C. Gardner}
\affiliation{Birck Nanotechnology Center, Purdue University, West Lafayette, Indiana 47907, USA}
\affiliation{School of Materials Engineering, Purdue University, West Lafayette, Indiana 47907, USA}
\author{M.\,J. Manfra}
\affiliation{Department of Physics and Astronomy, Purdue University, West Lafayette, Indiana 47907, USA}
\affiliation{Birck Nanotechnology Center, Purdue University, West Lafayette, Indiana 47907, USA}
\affiliation{School of Materials Engineering, Purdue University, West Lafayette, Indiana 47907, USA}
\affiliation{School of Electrical and Computer Engineering, Purdue University, West Lafayette, Indiana 47907, USA}
\affiliation{Station Q Purdue, Purdue University, West Lafayette, Indiana 47907, USA}
\received{\today}

\begin{abstract}
Our magnetotransport measurements of quantum Hall stripes in a high-quality GaAs quantum well in a slightly tilted magnetic field reveal that the orientation of stripes can be changed by temperature. 
Field-cooling and field-warming measurements, as well as observation of hysteresis at intermediate temperatures allow us to conclude that the observed temperature-induced reorientation of stripes is owing to the existence of two distinct minima in the symmetry-breaking potential. 
We also find that the native symmetry-breaking mechanism does not depend on temperature and that low-temperature magnetotransport data should be treated with caution as they do not necessarily reveal the true ground state, even in the absence of hysteresis.
\end{abstract}

\maketitle
Electronic nematic phases have been observed in a variety of systems \citep{fradkin:2010}, ranging from a clean two-dimensional(2D) electron gas (2DEG) formed in GaAs quantum wells \cite{lilly:1999a,du:1999,koulakov:1996,fogler:1996,fradkin:1999,fradkin:2000} to ruthenates \citep{borzi:2007}, high temperature superconductors \citep{daou:2010,chu:2010}, and heavy fermion systems \citep{okazaki:2011}. 
In a 2DEG in GaAs, the appearance of nematic phases (often referred to as stripes) is marked by the resistance minima (maxima) along the easy (hard) transport direction near half-integer values of the filling factor $\nu= n_eh/eB$, where $n_e$ is the electron density and $B$ is the magnetic field.
Stripe phases stem from the competition between (long-range) repulsive and (short-range) attractive components of Coulomb interaction and can be viewed as a unidirectional charge density wave consisting of stripe regions with either higher or lower integer filling factors.
While native stripes are usually aligned along the $\easy$ crystal direction \citep{note:0}, what exactly determines such preferred orientation remains a mystery \cite{sodemann:2013,koduvayur:2011,pollanen:2015}.

One way to learn about the native symmetry-breaking is to introduce some external mechanism which would cause a reorientation of stripes.
The oldest (and still the most popular) tool to reorient stripes is applying an in-plane magnetic field \cite{lilly:1999b,pan:1999,pan:2000,cooper:2001,jungwirth:1999,stanescu:2000,zhu:2002,cooper:2004,zhu:2009,pollanen:2015,shi:2016b,shi:2016c,shi:2016c,shi:gate}.
Other parameters which were manipulated to change the orientation of stripes include electron density \cite{zhu:2002,liu:2013}, filling factor within a given Landau level \cite{cooper:2004,shi:2016b}, electric current \citep{gores:2007}, mechanical strain \citep{koduvayur:2011}, symmetry of the quantum confinement containing 2DEG \citep{liu:2013}, capping layer thickness \citep{pollanen:2015}, and, most recently, weak periodic density modulation \citep{mueed:2016}.
It is also known that the orientation can be history-dependent and be governed by the direction of a density \cite{zhu:2002} or a magnetic field \cite{cooper:2004} sweep.
The observed hysteresis was attributed to a bi-directionality of the native symmetry breaking potential \cite{cooper:2004}.

In this Rapid Communication we report on magnetotransport measurements in a high-quality GaAs quantum well, subjected to a magnetic field which is tilted slightly away from the sample normal.
The weak in-plane magnetic field $\bip$ was introduced in order to reorient stripes only far away from half-filling \citep{shi:2016b} which is essential for our observations discussed below.
Our magnetotransport measurements revealed orthogonal orientations of quantum Hall stripes at different temperatures.
At a higher temperature ($T\approx 75$ mK) stripes near half-integer filling  factors are oriented along the native direction, $\y \equiv \easy$, as the magnitude of $\bip$ is not sufficient to induce reorientation. 
However, measurements at a base temperature ($T\approx 20$ mK) show a strong transport anisotropy indicative of a reorientation along the $\x \equiv \hard$ direction. 
Field-cooling/warming measurements and detection of hysteresis at intermediate temperatures suggest that the observed temperature-induced reorientation of stripes is due to the existence of two distinct minima in the symmetry-breaking potential. 
One important implication of our findings is that the low-temperature transport data do not necessarily reflect the equilibrium electron configuration even if no hysteresis is observed. 
Our findings also shed new light on the recently reported filling factor dependence of the native symmetry-breaking field \citep{shi:2016b}, e.g., that its observation calls for sufficiently high temperatures as it can be completely hidden in the low-temperature magnetotransport \citep{note:1}.
Finally, our data indicate a very weak, if any, temperature dependence of the native symmetry-breaking potential. 

Our sample is a $4\times4$ mm square cleaved from a symmetrically doped, 30 nm-wide GaAs/AlGaAs quantum well. 
Electron density and mobility were $n_e \approx 2.9 \times 10^{11}$ cm$^{-2}$ and $\mu \approx 1.6 \times 10^7$ cm$^2$/Vs, respectively \citep{manfra:2014}.
Eight indium contacts were fabricated at the corners and midsides of the sample. 
The longitudinal resistances, $\rxx$ and $\ryy$, were measured using standard four-terminal, low-frequency lock-in technique with the current (10 nA, 7 Hz) sent through the midside contacts and the voltage drop measured between the corner contacts.
All results were obtained with the magnetic field tilted by $\approx 6^\circ$ \citep{note:2} about the $\x$ axis with respect to the sample normal and swept at a rate of 0.1 T/min.

\begin{figure}[t]
\centering
\includegraphics{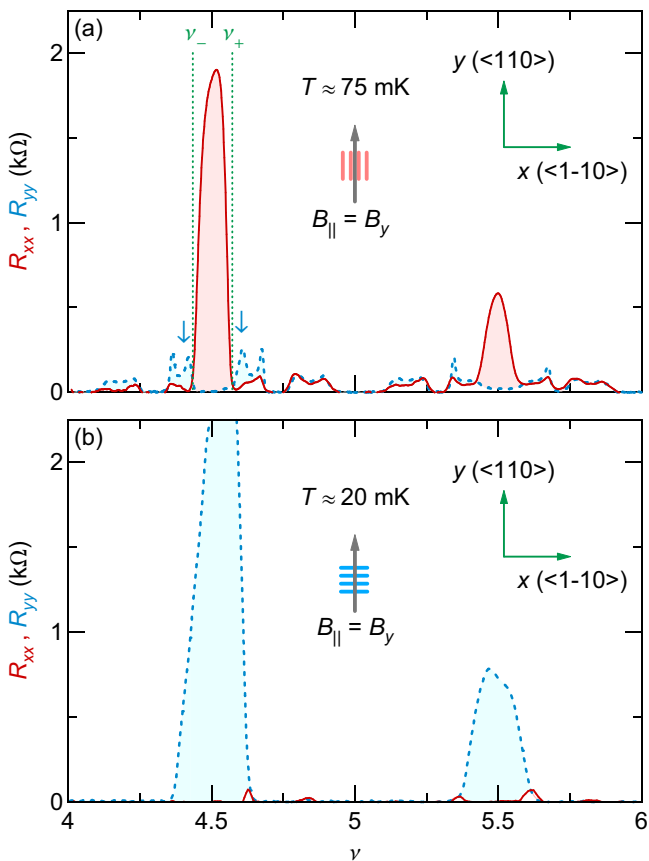}
\vspace{-0.1 in}
\caption{(Color online)
$\rxx$ (solid line) and $\ryy$ (dotted line) as a function of $\nu$ measured at (a) $T \approx 75$ mK and (b) $T \approx 20$ mK.
The magnetic field was tilted with respect to the sample normal by $\approx 6^\circ$.
Cartoons illustrate orientation of the stripes with respect to the direction of $\bip$ (thick arrow).
Dotted lines around $\nu = 9/2$ in panel (a) mark filling factors $\nu_+ \approx 4.57$ and $\nu_-\approx 4.43$ near which the anisotropy energy changes sign at this tilt angle.
}
\vspace{-0.1 in}
\label{fig1}
\end{figure}
In \rfig{fig1}(a) we present $\rxx$ (solid line) and $\ryy$ (dotted line) as a function of filling factor $\nu$ measured at $T \approx 75$ mK. 
The data demonstrate strong anisotropy (with $\rxx \gg \ryy$) near both $\nu = 9/2$ and $\nu = 11/2$, signaling that stripes at both filling factors are oriented along the native, $\y = \easy$, direction. 
Around fractional fillings, $\nu \approx 4.28$ and $\nu \approx 4.72$, both $\rxx$ and $\ryy$ reveal insulating states owing to the formation of bubble phases \citep{koulakov:1996,lilly:1999a,du:1999,eisenstein:2002,gores:2007,deng:2012a,deng:2012b}.
Closer examination of the data in \rfig{fig1}(a) around $\nu = 9/2$ reveals that further away from half-filling $\ryy \gg \rxx$ (marked by $\downarrow$). 
This observation indicates that stripes at these filling factors have, in fact, already undergone $\bip$-induced reorientation and are now aligned along the $\x$ direction.
This happens because stripes away from half-filling require a smaller $\bip$ to be reoriented, and, as a result, orientation of stripes within a given Landau level at a fixed tilt angle depends sensitively on the filling factor \citep{shi:2016b}.
As we show below, this filling factor dependence is responsible for the apparent reorientation of stripes seen in \rfig{fig1} (b) which we discuss next.

Like \rfig{fig1}(a), \rfig{fig1}(b) shows $\rxx$ (solid line) and $\ryy$ (dotted line) versus $\nu$ but measured at a lower temperature, $T \approx 20$ mK.
One observes that the bubble phases are now almost indistinguishable from the neighboring integer quantum Hall states.
The most striking change, however, is that now one finds $\rxx \ll \ryy$ near both $\nu = 9/2$ and $\nu = 11/2$, indicating alignment of stripes along the $\x$ direction. 

Since both data sets shown in \rfig{fig1} are acquired at the same tilt angle, one can argue that lowering the temperature is the sole cause for the observed reorientation.
If so, one could be tempted to conclude that, surprisingly, the native symmetry-breaking potential becomes \emph{weaker} at lower $T$ as it is now easier to be overcome by the $\bip$-induced anisotropy energy. 
However, as we show next, such a conclusion is premature.

First, as already mentioned, stripes in our sample have already been reoriented away from half-filling by $\bip$.
Second, we recall that the native stripes align along either $\y = \easy$ (most common scenario) or $\x = \hard$ orientation \cite{zhu:2002,cooper:2004,liu:2013,pollanen:2015} but never otherwise.
Moreover, there is strong evidence for the existence of two distinct minima in the native orientation potential along orthogonal directions \cite{cooper:2004}.
The anisotropy energy can be defined as a difference between these two minima, $E = E_x - E_y$.
While the orientation of stripes is usually decided by the sign of $E$ (i.e., $E > 0$ corresponds to stripes along the $\y$ direction), it can also be determined by the history.
The latter happens when (i) $E$ changes sign at some filling factor within a region where stripes form and (ii) the relaxation is slow enough for the system to relax from the higher minimum to the ground state during the magnetic field sweep.
Indeed, both \rref{zhu:2002} and \rref{cooper:2004} observed that when $\nu$ is swept from high to low (low to high) stripes near $\nu = 9/2$ align along the $\easy$ ($\hard$) direction.
Field-cooling \cite{cooper:2004} confirmed the existence of the filling factor at which $E$ changes sign.
Thus, observed hysteresis demonstrates that the orientation of stripes manifested in low-temperature measurements can be governed by the sign of $E$ preceding the degeneracy of the two minima.
In other words, no matter which orientation is realized first, stripes can get stuck in that orientation even though the sign of $E$ suggests an orthogonal one.

While magnetotransport at both $T \approx 75$ mK and $20$ mK showed negligible sensitivity to the field sweep direction, our experimental situation bears clear similarities to that of \rrefs{zhu:2002,cooper:2004}, albeit with one important difference.
Indeed, our higher temperature data clearly illustrate the existence of characteristic filling factors around $\nu = 9/2$, $\nu_+$ and $\nu_-$, separating stripes of orthogonal orientations [cf. dotted lines in \rfig{fig1}(a)]. 
The anisotropy energy is negative when $\nu_- < \nu < \nu_+$, vanishes at $\nu_+$ and $\nu_-$, and is positive further away from half-filling.
As a result, crossing either $\nu_-$ or $\nu_+$ leads to a sign change of $E$, very much like the situation of \rref{cooper:2004}.
The important distinction, however, is the existence of \emph{two} such filling factors in the filling factor range within which stripes form.
As we explain below, this fact plays a crucial role in our observation of the apparent $T$-induced reorientation illustrated in \rfig{fig1}.

\begin{figure}[t]
\includegraphics{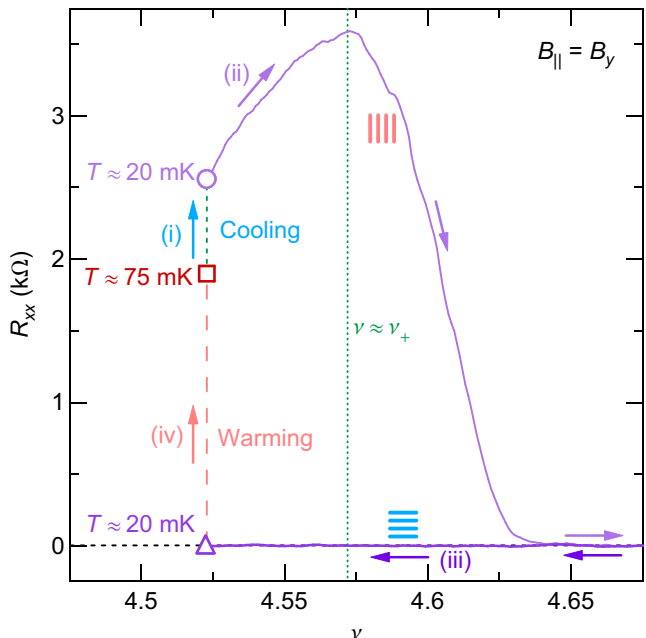}
\vspace{-0.15 in}
\caption{(Color online)
$\rxx$ vs. $\nu$ measured during (i) cooling (dotted line) from $T \approx 75$ mK (square) down to $T \approx 20$ mK (circle) at $\nu \approx 4.52$, (ii) down-field sweep at $T \approx 20$ mK (top curve), (iii) up-field sweep at $T \approx 20$ mK (bottom curve), and (iv) subsequent warming (dashed line) from $T \approx 20$ mK (triangle) to $T \approx 75$ mK (square) at $\nu \approx 4.52$.
Cartoons illustrate orientation of stripes.
The magnetic field was tilted with respect to the sample normal by $\approx 6^\circ$.
Vertical dotted line is drawn at $\nu = \nu_+ \approx 4.57$. 
}
\vspace{-0.15 in}
\label{fig2}
\end{figure}

We now argue that, in contrast to what the data in \rfig{fig1}(b) imply, stripes parallel to $\x$ do not reflect an equilibrium configuration of the system near $\nu = 9/2$ and 11/2. 
To this end, we discuss additional experiments the results of which are summarized in \rfig{fig2}.
We cool our sample at $\nu \approx 4.52$ \citep{note:4} starting from $T \approx 75$ mK (square) down to $T \approx 20$ mK (circle) and observe an increase in $\rxx$ (dotted line), suggesting that stripes stay parallel to $\y$.
We next sweep the magnetic field away from the stripe region (upper curve), return back to $\nu \approx 4.52$ (lower curve), and find that $\rxx \approx 0$, i.e. that stripes are now parallel to $\x$. 
Subsequent warming (dashed line) leads to an increase of $\rxx$ signaling reorientation of stripes back to along the $\y$ direction.
We also note that neither up-sweep nor down-sweep reveals any sensitivity of orientation to the filling factor, in contrast to the higher-$T$ data of \rfig{fig1}(a).

The data in \rfig{fig2} suggest that when $\nu_-<\nu<\nu_+$, the equilibrium configuration corresponds to stripes along the $\y$ direction, in accord with \rfig{fig1}(a). 
When the system is cooled at $\nu$ within the range $\nu_-<\nu<\nu_+$, the orientation does not change.
However, after cooling at $\nu$ outside this range of filling factors, magnetoresistance reflects the orthogonal orientation as shown in \rfig{fig1}(b).
This happens because before entering the range $\nu_-<\nu<\nu_+$ ($E > 0$), one first needs to traverse the regime favoring stripes along the $\x$ direction ($E < 0$) which occurs at both $\nu < \nu_-$ and $\nu > \nu_+$. 
At low $T$, the relaxation time is long enough to preserve the initial orientation for the duration of the field sweep.
Indeed, crossing the dotted line drawn at $\nu  = \nu_+$ from either side does not trigger the reorientation of stripes.
As mentioned above, the unique feature of our experiment is the existence of two characteristic filling factors at which $E$ changes sign.
It is this feature which leads to persistent observation of stripes of metastable orientation without hysteresis at low temperatures, as shown in \rfig{fig1}(b).

\begin{figure}[t]
\includegraphics{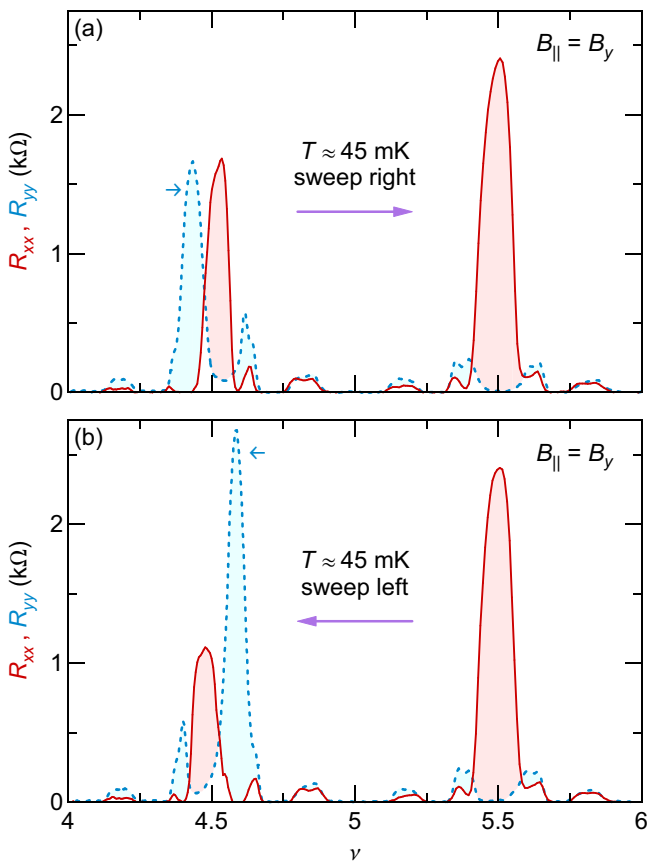}
\vspace{-0.15 in}
\caption{(Color online)
$\rxx$ (solid line) and $\ryy$ (dotted line) as a function of $\nu$ measured in (a) down-sweep and (b) up-sweep at $T \approx 45$ mK.
The magnetic field was tilted with respect to the sample normal by $\approx 6^\circ$.
}
\vspace{-0.15 in}
\label{fig3}
\end{figure}

While no hysteresis is seen at either $T \approx 20$ mk or $T \approx 75$ mK (cf. \rfig{fig1}), we do indeed observe sensitivity to the field sweep direction at intermediate temperatures.
In \rfig{fig3} we present $\rxx$ (solid line) and $\ryy$ (dotted line) measured in (a) up-sweep and (b) down-sweep at $T \approx 45$ mK.
One immediately observes that the data near $\nu = 9/2$ reveal dramatic differences depending on the magnetic field sweep direction.
Indeed, when the region with stripes is approached from either higher or lower $\nu$, the first anisotropic phase characterized by $E<0$ (stripes parallel to $\x$) is marked by a stronger resistance anisotropy [cf.\,$\rightarrow$ in (a), $\leftarrow$ in (b)] than its counterpart which occurs on the opposite side of half-filling. 
In fact, the anisotropy away from half-filling can be as strong or even stronger than the anisotropy at exactly $\nu = 9/2$, in sharp contrast with the data in \rfig{fig1}(a).
This behavior manifests that a relaxation time at this $T$ is comparable to the sweep time of about a minute \citep{note:3}. 
At higher $T$, the relaxation becomes considerably faster which results in the data presented in \rfig{fig1}(a) showing the equilibrium orientation of stripes with no hysteresis.
At lower $T$, the relaxation is slower, and the transport data show the metastable orientation which is set by the sign of $E$ away from half-filling.

Further examination of the data in \rfig{fig3} shows that, in contrast to $\nu = 9/2$, the magnetoresistance near $\nu =11/2$ exhibits virtually no dependence on the sweep direction.   
The absence of hysteresis near $\nu = 11/2$ indicates that the relaxation time at this $T$ is short enough for the transport to always reflect the true ground state.
Faster relaxation near $\nu = 11/2$ can be attributed to a larger $|E|$ at this filling factor compared to that at $\nu = 9/2$.
Indeed, $\bip$-induced reorientation requires smaller $\bip$ at $\nu =9/2$ than at $\nu = 11/2$ in our sample \citep{shi:2016c}.
Furthermore, a faster relaxation near $\nu = 11/2$ can also explain a lower resistance anisotropy at $T \approx 20$ mK shown in \rfig{fig1}(b) compared to $T \approx 45$ mK data shown in \rfig{fig3}. 
Here, although the majority of the domains remain in a metastable orientation during the magnetic field sweep, some of the domains might have managed to relax to the ground state, thus lowering the resistance anisotropy at half-filling.

Finally, our findings allow us to revisit the conclusion of \rref{shi:2016b}, namely, that the magnitude of the native anisotropy energy quickly decays as the filling factor deviates from half-filling.
Having confirmed the existence of the two distinct minima in the native orientation potential in our sample, we can now conclude that stripes near (away from) half-filling are coupled more strongly to the minimum of the native potential which favors stripes along the $\easy$ ($\hard$) direction.
This finding should be taken into account by future proposals concerning the nature of the native symmetry breaking field.

In summary, magnetoresistance measurements in a high-quality GaAs quantum well with the magnetic field tilted slightly away from the sample normal revealed stripes along the $\easy$ at $T \approx 75$ mK and along $\hard$ at $T \approx 20$ mK.
Sample cooling at a fixed filling factor near $\nu = 9/2$, followed by magnetic field sweeps and warming at the same filling factor, as well as the observation of hysteresis at intermediate temperatures, let us conclude that, in contrast to $\rxx \ll \ryy$ detected in magnetotransport at $T \approx 20$ mK, stripes along $\easy$ crystal direction represent the true ground state. 
These findings demonstrate that the low-temperature magnetoresistance, which is routinely used to determine orientation of stripes, can be misleading as, even in the absence of hysteresis, it does not necessarily reflect the ground state. 
In addition, a recently reported filling factor dependence of reorientation of stripes \citep{shi:2016b} can be completely hidden in the low-temperature magnetotransport.
Finally, since our data were obtained on the verge of reorientation ($E \approx 0$), we can also conclude that the native symmetry-breaking potential depends on temperature only weakly, if at all, in our experiment. 
While in the present work we have employed $\bip$ along $\y = \easy$ direction to introduce $\nu_+$ and $\nu_-$, similar observations can be expected under orthogonal orientation of $\bip$, as it can also induce reorientation of stripes \citep{shi:2016c}.

\begin{acknowledgments}
We thank J. Geurs for assistance with the experimental setup and I. Dmitriev for commenting on the manuscript.
The work at Minnesota (Purdue) was supported by the U.S. Department of Energy, Office of Science, Basic Energy Sciences, under Award \# ER 46640-SC0002567 (DE-SC0006671).
Q.S. acknowledges University of Minnesota Doctoral Dissertation Fellowship.
\end{acknowledgments}

\small{$^\#$Current address: QuTech and Kavli Institute of Nano\,-science, Delft Technical University, 2600 GA Delft, The Netherlands.}


\begin{thebibliography}{38}
\bibitem[{not({\natexlab{d}})}]{note:4}
\bibinfo{note}{$\rxx$ near $\nu = 9/2$ is slightly asymmetric and peaks at $\nu
  \approx 4.52$ in \rfig{fig1} (a). At low $T$ the peak moves to even higher
  $\nu$ which might account for the initial rise of $\rxx$ observed in the
  up-sweep (ii) of \rfig{fig2}.}
\end{thebibliography}

\end{document}